\documentclass[journal=nalefd,manuscript=article]{achemso}
\usepackage[version=3]{mhchem} % Formula subscripts using \ce{}
\setkeys{acs}{articletitle = true}
\setkeys{acs}{keywords = true}
\SectionsOn

\usepackage{mathtools} 
\usepackage{hyperref}
\hypersetup{
    colorlinks=true,
    urlcolor  = magenta,
    citecolor=blue, % cite
    linkcolor=magenta % ref
}
\usepackage{ulem}
\usepackage{tikz}

\makeatletter
\newcommand{\newsout}[1]{%
  {\bgroup
   \ULdepth=.5ex % position the line in the middle
   \def\ULthickness{0.9pt}%
   \markoverwith{\textcolor{red}{\rule[.5ex]{2pt}{\ULthickness}}}%
   \ULon{#1}%
   \egroup}%
}
\usepackage{amsmath}
 \usepackage{amssymb,stackengine,graphicx}
\usepackage{xcolor}
\usepackage[mathscr]{euscript}
\usepackage{orcidlink}
\usepackage{wrapfig}
\usepackage{makecell}
   \author{S. Pal\,\orcidlink{0009-0003-6356-3409}\,}
    \email{palsubhojit429@gmail.com}
    \affiliation{\rm Dipartimento di Fisica e Chimica-Emilio Segr\`e, \href{https://www.unipa.it/target/international-students/en/about/the-university/}{Universit\`a degli Studi di Palermo}, Via Archirafi 36, 90123 Palermo, Italy}

    \author{L. M. Woods\,\orcidlink{0000-0002-9872-1847}\,}
\affiliation{Department of Physics, \href{https://www.usf.edu}{University of South Florida}, Tampa, FL, 33620, USA}

\author{C. Persson\,\orcidlink{0000-0002-9050-5445}\,}
\affiliation{Department of Materials Science and Engineering, \href{https://www.kth.se/en}{KTH Royal Institute of Technology}, SE-100 44 Stockholm, Sweden}

\author{I. Brevik\,\orcidlink{0000-0002-9793-8278}\,}
\affiliation{Department of Energy and Process Engineering, \href{https://www.ntnu.edu/}{Norwegian University of Science and Technology}, NO-7491 Trondheim, Norway}

 \author{U. De Giovannini\,\orcidlink{0000-0002-4899-1304}\,}
 \email{umberto.degiovannini@unipa.it} 
  \affiliation{\rm Dipartimento di Fisica e Chimica-Emilio Segr\`e, \href{https://www.unipa.it/target/international-students/en/about/the-university/}{Universit\`a degli Studi di Palermo}, Via Archirafi 36, 90123 Palermo, Italy}
    \alsoaffiliation{\href{https://www.mpsd.mpg.de/person/44457/2736}{Max Planck Institute for the Structure and Dynamics of Matter}, Luruper Chaussee 149, Hamburg, Germany}

    \author{M. Bostr{\"o}m\,\orcidlink{0000-0001-5111-4049}\,}
    \email{mathias.bostrom@ensemble3.eu}
     \affiliation{\href{https://ensemble3.eu/}{Centre of Excellence ENSEMBLE3} Sp. z o. o., Wolczynska Str. 133, 01-919, Warsaw, Poland}
    \alsoaffiliation{Chemical and Biological Systems Simulation Lab, \href{https://cent.uw.edu.pl/}{Centre of New Technologies, University of Warsaw}, Banacha 2C, 02-097 Warsaw, Poland}

\title[Nanorod Pair Complexes Manipulated via Magnetic Casimir Forces]{Nanorod Pair Complexes Manipulated via Magnetic Casimir Forces}

\begin{document}

\newpage

 \begin{abstract}

% \noindent
% \makebox[\textwidth]{\textbf{\Large Abstract}}
% % \vspace{2em}
% \begin{wrapfigure}{r}{0.48\textwidth}
%     \vspace{-2pt} % adjust if needed
%     \centering
%     \includegraphics[width=0.48\textwidth]{sections/TOC.pdf}
%     \vspace{-20pt} % adjust if needed
% \end{wrapfigure}
% \noindent
Controlling nanoscale interactions to suppress aggregation from short-range attractive forces is a key problem in nanoengineering. Here, we demonstrate a route to modulate Casmir-Lifshitz interactions between anisotropic nanoparticles with the magnetic fluids. By semi-classical quantum electrodynamics, we study ground state dispersion forces for cylindrical dielectric nanorods made of polystyrene (PS), and zinc oxide (ZnO) embedded in toluene-based host media with gold-coated magnetite nanoparticles and also predict magnetic contributions to the fully retarded excited state interaction. The variation in magnetic permeability enables tuning between repulsive and attractive interaction and measurable magnetic Casimir traps are predicted between a pair of ZnO–PS nanoparticles whose equilibrium position can be modulated over an order of magnitude with a small variation in the size of the magnetite nanoparticle. This provides an alternative magnetic Casimir-effect pathway to reversibly tune quantum electromagnetic forces at the nanoscale for assembly and enhancement of colloidal stability.
\end{abstract}

% In nanoengineering it is important to avoid attraction from short-range forces that can cause the clustering of building blocks. In this work, a method is proposed to go beyond systems that produce pure attraction or pure repulsion. 
% The dispersion interaction between pairs of thin elongated nanoparticles in a magnetic fluid is explored, demonstrating the importance of the magnetic contributions in the long-range and high-temperature limits for excited and ground-state dispersion interactions.
% In practical calculations, we consider two different cylindrical nanoparticles: polystyrene (PS), Teflon (PTFE), and zinc oxide (ZnO). We focus on interactions taking place in fluids made from toluene mixed with gold-coated magnetite particles. Using a pair of ZnO and PS nanoparticles we predict stable equilibria from magnetic Casimir trapping. Notably, the trapping distances are tuned more than one order of magnitude entirely by changes in attractive zero-frequency magnetic contributions to the Casimir force. This can be achieved via changes in the average diameter for the magnetite particles while keeping volume fractions of magnetite fixed. }
% \end{abstract}
% \date{\today}

\newpage
\maketitle
% \section{Introduction}

% \input{sections/introduction}

% \input{sections/introduction_jeet}
% \input{sections/preliminary}
% \input{sections/theory}
% \input{sections/excited_states}
% \input{sections/results}
% \input{sections/discussion_conclusion}

What if the very surrounding media in space itself could be engineered into a tool for nano-manipulation? For instance, can we take the background media as a magnetic  fluid, in which  electromagnetic fluctuations exert forces that can pull, push  or most interestingly trap matter  at the nanoscale level? Such fluctuation-induced interactions, which are   referred to as van   der Waals\,\cite{wang1928problem,Lond} (vdW) and Casimir-Lifshitz (CL)\,\cite{casimir1948-1,Dzya} forces together,  typically act to bring microscopic bodies into contact, leading to  unwanted self-aggregation in nanofabrication and soft-matter systems\,\cite{ParsegianNinham1969,dobson2006asymptotics,Munday2009,esteso2015nanolevitation,WoodsRevModPhys.88.045003,zhao2019stable,Woods2020,esteso2022effect,MundayTorque2022,MundayReview2024,Woods2024}.
An important practical issue in nano and micro-devices is to overcome close contact attractions that causes building blocks to cluster together and an experimental solution is to engineer systems where the interaction turns repulsive at small distances\,\cite{Munday2009,esteso2015nanolevitation,WoodsRevModPhys.88.045003,zhao2019stable,Woods2020,esteso2022effect,MundayReview2024,WilliamsonJPCL2025}
% . This is a hot topic and a proposed solution to this problem is engineering systems whose interaction at short distances turns repulsive\,\cite{MundayReview2024}.
Beyond simply mitigating stiction, these fluctuations could also be harnessed to produce stable and controllable interaction potentials.
{Fundamental investigations of the influence on CL forces by a magnetic surrounding media  have been largely restricted to simple geometries\,\cite{Richmond_1971,klimchitskaya2019impact,VelichkoKlimchitskaya2020}. Exploiting such a foundation, we have recently expanded the theoretical approach by including the contributions of transverse magnetic (TM) and transverse electric (TE) modes within a model described by the proximity force theorem\,\cite{Derjaguin1934,BLOCKI198153} in such a way that both energy contributions due to TM and TE modes would partially act as candidates to enable quantum trapping between two quasi-planar surfaces\,\cite{pal2025trapping}.}

% Fundamental explorations to study the impact of magnetic media on CL forces have been carried out for simple geometries\,\cite{Richmond_1971,klimchitskaya2019impact,VelichkoKlimchitskaya2020}.  In an expansion of previous work, using the proximity force theorem\,\cite{Derjaguin1934,BLOCKI198153}, we recently found that combined energy contributions from transverse magnetic (TM) and transverse electric (TE) terms potentially can control quantum trapping between quasi-planar surfaces\,\cite{pal2025trapping}.

{In this letter, we go beyond these simple geometries and study an experimentally feasible setup which comprises of pairs of cylindrical nanorods made of polystyrene (PS) and zinc oxide (ZnO). The use of the full cylindrical geometry permits a more realistic description of nanoscale behavior, albeit at the price of considerable added theoretical complexity. This stimulates further development of analytical and numerical studies of Lifshitz and vdW interaction between elongated nanoparticles\,\cite{Richmond1972,mitchell1973van,Davies1973,NylandBrevik1994,SubhojitPCCP2024,PhysRevE.111.015403}.}
% The explicit experimentally relevant system we consider in the current work are pairs of cylindrical polystyrene (PS), Teflon (PTFE) and zinc oxide (ZnO) nanorods. Going to the actual geometry of pairs of cylindrical nanoparticles will lead us to substantially improved modeling, but also a much more complex theory.
% This requires us to expand on the existing works\,\cite{Richmond1972,mitchell1973van,Davies1973,NylandBrevik1994,SubhojitPCCP2024,PhysRevE.111.015403} on vdW and CL interactions between pairs of cylindrical elongated nanoparticles.  
One of us previously derived a few asymptotic limits for this kind of a system\,\cite{NylandBrevik1994}. Here, we examine the \textit{magnetic contributions}  associated with a surrounding ferrofluid for the long-range and high-temperature limits of ground-state dispersion interactions between cylindrical nanoparticles.  We also predict a new kind of magnetic contribution to the retarded excited-state interaction. In an illustrative example, we study the CL interaction between a ZnO nanorod and a PS nanorod suspended in a toluene and gold-coated magnetite nanoparticle (5-7\%) fluid mixtures. {In the case of two unequal nanorods, we find that subtle change of nanoparticle concentration into the ferrofluid can induce a transition from short-range repulsion to long-range attraction between them. It is the fluid’s magnetic permeability and its dependence upon the magnetite nanoparticle size that dictates both interaction strength and sign, in contrast to the fluid dielectric function which remains nearly constant. Such a device could enable controlled, magnetic-induced entropic CL trapping}.
% Notably, for two unequal elongated nanoparticles, the short-range repulsion to long-range attraction transition for magnetic CL interaction can be tuned using small changes in the nanoparticle mixing ratio in the fluid system. The interaction depends strongly on the permeability of the fluid via the diameter of the magnetite particles. The fluid dielectric function is kept stable while the magnetic permeability is manipulated to control magnetically induced entropic CL trapping. 
Early studies of CL interactions between magnetic fluida\,\cite{klimchitskaya2019impact,VelichkoKlimchitskaya2020} were mainly concentrated on the observation of a repulsive force by changing the optical properties of the medium. By contrast, we have implemented here a way in which this \textit{zero-frequency, purely entropically driven Casimir trapping}  is realized exclusively by tuning the fluid’s magnetic permittivity, an effect not investigated earlier.
% Previous works on CL forces between planar surfaces across magnetic fluids\,\cite{klimchitskaya2019impact,VelichkoKlimchitskaya2020} focused on achieving repulsion at all relevant length scales and the result was mainly impacted by varying the optical properties in the fluid mixture. However, purely zero-frequency entropically driven CL trapping entirely controlled via changes in magnetic permeability of the fluid media has not been previously explored. 

To lay foundations for such a framework, we develop, within the semi-classical quantum electrodynamics, the general theory for ground and excited state dispersion interactions for dielectric cylindrical nanorods embedded in a magnetic fluid. In the subsequent sections, we present analytical results for the excited state interaction between thin cylinders and the corresponding free energy, followed by numerical analysis of the ground-state forces. We briefly describe the essential optical and magnetic properties of the interacting nanorods and background fluid in the main text and at length in the Supplementary Information. Lastly we demonstrate here that the retardation substantially  diminishes the overall interaction  strength as well as amplifies the relative weight of magnetic response from the long-range, classical entropic contributions.
%in the long-range, classical, entropic case.}

Here, we thus propose a novel experimentally feasible approach to manipulate nanorod pair complexes by suspending them in a magnetically diluted fluid, where the interplay between repulsive non-magnetic and attractive magnetic contributions to the CL force can be continuously tuned to form  magnetic CL traps appearing as equilibria whose strength and range derive not from mechanical considerations but from spectral properties of quantum electromagnetic fluctuations. 

% \section{Theory}
\begin{figure}[!h]
    \centering
    \includegraphics[width=0.8\linewidth]{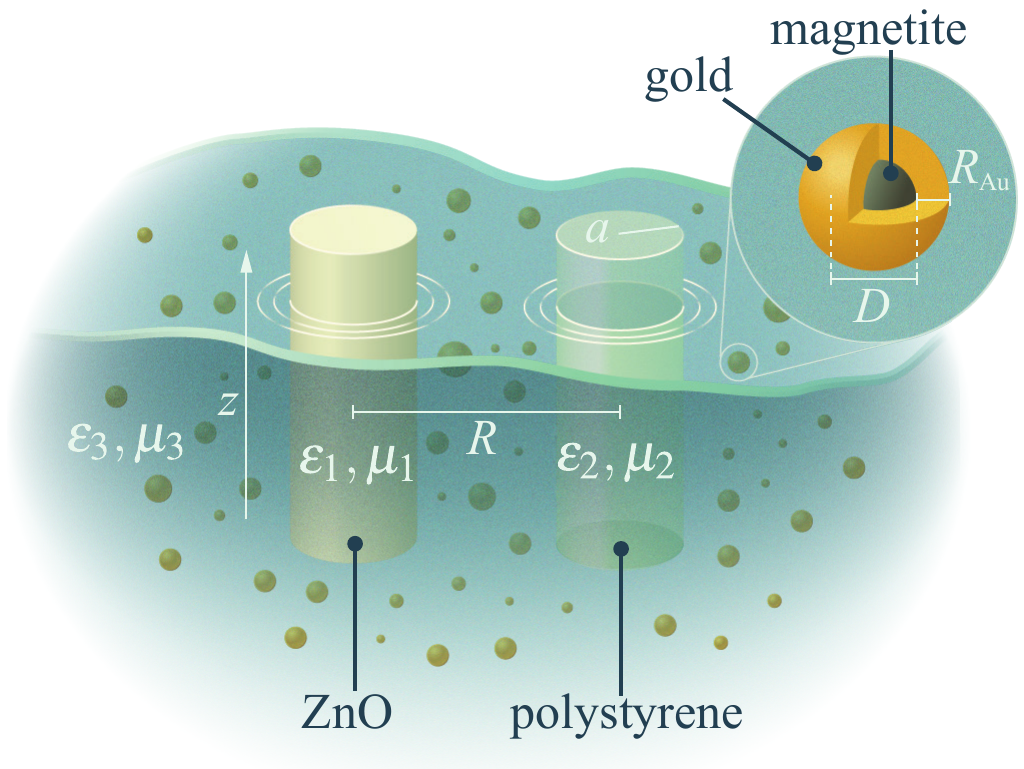}
    \caption{Color online: Two cylindrical nanorods, one composed of ZnO with permittivity $\varepsilon_1$ and permeability $\mu_1$, and the other of PS with permittivity $\varepsilon_2$ and permeability $\mu_2$, each having an identical radius $a$, are immersed in a magnetic liquid. The magnetic liquid is composed of toluene and it contains dispersed  gold coated (thickness $R_{\rm Au}$) magnetite nanoparticles of diameter $D$ and volume fraction $\phi$ and it has effective permittivity $\varepsilon_3$ and permeability $\mu_3$. The nanorods are separated by a center-to-center distance $R$.}
    \label{fig:scheme}
\end{figure}
We consider a system of two parallel cylindrical objects of radii $a$ with dielectric constant and magnetic permittivity, ($\varepsilon_1$, $\mu_1$) and ($\varepsilon_2$, $\mu_2$), respectively, embedded in a ferrofluid of dielectric constant $\varepsilon_3$ and permittivity $\mu_3$ shown in Fig.\,(\ref{fig:scheme}). The cylinders are separated by a center-to-center distance $R$. We assume that the cylinders are infinitely long in the $z$ direction, and we will work in the quasi-static limit  where the wavelength of light is much larger than any other length scale in the problem. The dispersion interaction in this system is obtained using scattering approach \cite{NylandBrevik1994}, where the interaction energy per unit length (for a cylinder with length $L$) in finite temperature $T$  is  (an explicit derivation is given in Sec.\,(S.2) of the Supplementary Information),
\begin{equation}
    G(a,R, T) = \frac{k_{B}T}{\pi}  \sum_{p=0}^{\infty}{}' \int_{0}^{\infty} dk\,\ln \mathscr{D}(i \xi_p) 
\end{equation}
where the prime in the summation means that the term for which $p=0$ must be weighted by $1/2$. The above expression is given in Matsubara frequencies  $\xi_p=2 \pi k_{B} T p/\hbar$, where $p$ is an integer. Here, $\mathscr{D}(i \xi_p)$  is expressed as the determinant of the scattering matrix of the interacting cylinders. In general, such a matrix is infinite, but significant simplifications can be made in the ``thin cylinder'' approximation where the separation distance between the cylinders is much larger than their radii ($R \gg a$). In this case, terms containing Bessel functions\cite{watson1958} with $n\geq 2$ make negligible contributions (details in Sec.\,(S.2) in the  Supplementary Information) as they rapidly decay with $R$. Thus, we  find that (see Sec.\,(S.2) of the Supplementary Information),
\begin{equation}
    \begin{aligned}
  \ln\mathscr{D}(i \xi_p) = f_0 K^{2}_{0}(\nu_3 R) +  f_1 K^{2}_{1}(\nu_3 R) + f_2 \Big(K^{2}_{0}(\nu_3 R) + K^{2}_{2}(\nu_3 R) \Big)
    \end{aligned}
\end{equation}
where $K_{0}, K_{1}$ and $K_{2}$  are modified Bessel functions of zeroth, first and second orders respectively,  $\nu^2_{\alpha}(i \xi_p) =  \Big( \frac{\xi^2_p}{c^2} \mu_{\alpha}(i\xi_p) \varepsilon_{\alpha}(i\xi_p) + k^2\Big)$ for $\alpha = 1,2,3$, and the coefficients are as follows:
\begin{equation}
    f_0 = - \frac{a^4}{4} \Bigg[ \frac{(\varepsilon_3 -\varepsilon_1)(\varepsilon_3 -\varepsilon_2)}{\varepsilon^2_3 } + \frac{(\mu_3-\mu_1)(\mu_3-\mu_2)}{\mu^2_3}\Bigg] \nu^4_3
\end{equation}
\begin{equation}
    \begin{aligned}
        f_1 = -a^4 \Bigg\{ \frac{\nu^2_1 \nu^4_3}{(\varepsilon_3 +\varepsilon_1)(\mu_3+\mu_1)}\Bigg[ \frac{(\varepsilon_3 - \varepsilon_2)}{\varepsilon_3}\Bigg(\frac{\varepsilon_3 \mu_3}{\nu^2_3} -\frac{\varepsilon_1 (\mu_1  +\mu_3) + \varepsilon_3 (\mu_3 - \mu_1)}{2\nu^2_1}\Bigg) +  \frac{(\mu_3 -\mu_2)}{\mu_3} \\ \Bigg(\frac{\varepsilon_3 \mu_3}{\nu^2_3} -\frac{\varepsilon_3(\mu_1  +\mu_3) + \varepsilon_1 (\mu_1 - \mu_3)}{2\nu^2_1}\Bigg)\Bigg] +  \frac{\nu^2_2 \nu^4_3}{(\varepsilon_3 +\varepsilon_2)(\mu_3+\mu_2)}\Bigg[ \frac{(\varepsilon_3 - \varepsilon_1)}{\varepsilon_3}\\\Bigg(\frac{\varepsilon_3 \mu_3}{\nu^2_3}  -\frac{\varepsilon_2 (\mu_2  +\mu_3) + \varepsilon_3 (\mu_3 - \mu_2)}{2\nu^2_2}\Bigg)  + \frac{(\mu_3 -\mu_1)}{\mu_3}\Bigg(\frac{\varepsilon_3 \mu_3}{\nu^2_3} -\frac{\varepsilon_3 (\mu_2  +\mu_3) + \varepsilon_2 (\mu_2 - \mu_3)}{2\nu^2_2}\Bigg)\Bigg] \Bigg\}
    \end{aligned}
\end{equation}

\begin{equation}
    \begin{aligned}
        f_2 = - \frac{a^4 \nu^2_1 \nu^2_2 \nu^4_3}{(\varepsilon_3 + \varepsilon_1)(\varepsilon_3+ \varepsilon_2)(\mu_3 + \mu_1)(\mu_3 + \mu_2)}\Bigg[ \Bigg(\frac{\varepsilon_3 \mu_3}{\nu^2_3} -\frac{\varepsilon_1 (\mu_1  +\mu_3) + \varepsilon_3 (\mu_3 - \mu_1)}{2\nu^2_1}\Bigg) \\ 
        \times \Bigg(\frac{\varepsilon_3 \mu_3}{\nu^2_3}  -\frac{\varepsilon_2 (\mu_2  +\mu_3) + \varepsilon_3 (\mu_3 - \mu_2)}{2\nu^2_2}\Bigg) + \Bigg( \frac{\varepsilon_3 \mu_3}{\nu^2_3} - \frac{\varepsilon_3 \mu_3}{\nu^2_1}\Bigg)\Bigg( \frac{\varepsilon_3 \mu_3}{\nu^2_3} - \frac{\varepsilon_2\mu_2}{\nu^2_2}\Bigg)\\ + \Bigg( \frac{\varepsilon_3 \mu_3}{\nu^2_3} - \frac{\varepsilon_3 \mu_3}{\nu^2_2}\Bigg)\Bigg( \frac{\varepsilon_3 \mu_3}{\nu^2_3} - \frac{\varepsilon_1\mu_1}{\nu^2_1}\Bigg) + \Bigg(\frac{\varepsilon_3 \mu_3}{\nu^2_3} -\frac{\varepsilon_3(\mu_1  +\mu_3) + \varepsilon_1 (\mu_1 - \mu_3)}{2\nu^2_1}\Bigg)\times \\ 
        \Bigg(\frac{\varepsilon_3 \mu_3}{\nu^2_3} -\frac{\varepsilon_3 (\mu_2  +\mu_3) + \varepsilon_2 (\mu_2 - \mu_3)}{2\nu^2_2}\Bigg)\Bigg]
    \end{aligned}
\end{equation}
 where  $c$ is the speed of light in vacuum.

Here we are also interested  if a dispersion interaction in excited state could be realized between two equal cylindrical nanoparticles. The interaction energy is due to the separation dependence of the excited energy, such that $E^{\text{res}}(R)= \hbar [\omega_{r} (R)-\omega_{r} (\infty)]$, where $\omega_{r} (R)$
 represents the resonance frequency at interatomic separation 
$R$. The interaction energy of the excited state is \cite{McLachlan,Bostrom1}, $E^{\text{res}}(R)=s^2 T(R|\omega_j)\propto R^{-3},$
where $s$ is the magnitude of the transition dipole moment of the excitation and  $T(R|\omega_j)$  is the dipole–dipole coupling tensor (from the electromagnetic Green’s function) at frequency $\omega_j$ which accounts for how field from one oscillating dipole  influences another at distance $R$ and it typically scales as $R^{-3}$ in the non-retarded limit. This can be compared with the ground state van der Waals interaction 
which decays with distance as $R^{-6}$\,\cite{Parsegian2006}. %(\textcolor{red}{some reference to London vdW with a Ref is needed here, I think}) 
%This is the correct textbook result for low-temperature excited state interaction energy between  two identical atoms in an excited state in the non-retarded limit (\textcolor{red}{do not understand this sentence, maybe rephrase and give a Ref?}).  

% \subsection{Excited state interaction between thin cylinders in a magnetic fluid}
This analogy implies that two identical cylindrical nanoparticles can similarly form a long lasting antisymmetric excited state by virtue of a long-range resonant dispersion interaction. A corresponding expression for the retarded excited-state interaction energy, obtained semiclassically in Sec.\,(S.3) of the Supplementary Information, reads 

\begin{equation}
\begin{aligned}
    G^{\text{res}}(a,R,T) & = -\frac{2k_{B}T}{\pi} \sum_{p=0}^{\infty}{}'\int_{0}^{\infty} dk\,(t_{0} + t_{1} + t_2 + t_3 )\,K_{0}(\nu_3 R)
    \end{aligned}
    \label{s63}
\end{equation}

where 
\begin{equation}
    \begin{aligned}
        t_0 = \Bigg[\frac{\varepsilon_{3} -  \mu_{1}}{2\,\varepsilon_{3}}\Bigg] a^2 \nu^2_{3}; \quad \quad t_1 = \Bigg[\frac{\mu_{3} -  \varepsilon_{1}}{2\,\mu_{3}}\Bigg] a^2 \nu^2_{3}\\
        t_3 = \Bigg[-\frac{\varepsilon_1 (\mu_1  +\mu_3) + \varepsilon_3 (\mu_3 - \mu_1)}{\nu^2_1} + \frac{2 \varepsilon_3 \mu_3}{ \nu^2_3}\Bigg] \frac{a^2\nu^2_1\nu^2_3}{ (\varepsilon_1 + \varepsilon_3) (\mu_1 + \mu_3)}\\
         t_4 = \Bigg[-\frac{\varepsilon_3 (\mu_1  +\mu_3) + \varepsilon_1 (\mu_1-\mu_3 )}{\nu^2_1} + \frac{2 \varepsilon_3 \mu_3}{ \nu^2_3}\Bigg] \frac{a^2\nu^2_1\nu^2_3}{ (\varepsilon_1 + \varepsilon_3) (\mu_1 + \mu_3)}
    \end{aligned}
\end{equation}
% \begin{equation}
% \begin{aligned}
%     G^{\text{res}}_{p=0}(a,R,T)\Big|_{\text{non-magnetic}} &= -\frac{k_{B}T a^2}{\pi} \Bigg[\frac{\varepsilon_{3}(0) -  \varepsilon_{1}(0)}{2\,\varepsilon_{3}(0)}\Bigg]  \int_{0}^{\infty} dk \, k^2 \, K_{0}(kR)\\
%     &= -\frac{k_{B}T a^2}{4 R^3} \Bigg[\frac{\varepsilon_{3}(0) -  \varepsilon_{1}(0)}{\varepsilon_{3}(0)}\Bigg]  
%     \end{aligned}
%      \label{Resonance_nonmagnetic}
% \end{equation}

% \begin{equation}
% \begin{aligned}
%     G^{\text{res}}_{p=0}(a,R,T)\Big|_{\text{magnetic}} &= -\frac{k_{B}T a^2}{\pi} \Bigg[\frac{\mu_{3}(0) -  \mu_{1}(0)}{2\,\mu_{3}(0)}\Bigg]  \int_{0}^{\infty} dk \, k^2 \, K_{0}(kR)\\
%     &= -\frac{k_{B}T a^2}{4 R^3} \Bigg[\frac{\mu_{3}(0) -  \mu_{1}(0)}{\mu_{3}(0)}\Bigg]  
%     \end{aligned}
%     \label{Resonance_magnetic}
% \end{equation}
% \end{widetext}
The entropic contribution ($p=0$) suggests that Eq.\,(\ref{s63}) provides $R^{-3}$ dependence that is much longer ranged than the ground state interaction between two equal thin cylindrical nanoparticles that is proportional to $R^{-5}$\,\cite{langbein1972van}. It gives both magnetic and non-magnetic contributions. Notably, this previously neglected magnetic contribution in Eq.\,(\ref{s63}) has the same power law as its non-magnetic contribution.  However, the significance of this contribution should be judged by the low temperature/short distance non-magnetic excited state interaction. 
 % This leads to an interaction decaying as $R^{-3}$, as shown in \textcolor{red}{Eq.??}  
The non-retarded excited state interaction becomes attenuated by retardation due to the finite velocity of light.  
In the retarded limit additional effects potentially impact the interaction via real photon exchange. 
%\mb{[Mathias: This sentence should be removed:] However, such effects are expected to be small when the interaction is within a fluid} (\textcolor{red}{good to have a reference here}).
Ultimately, in the limit of large distances or high temperatures, the interaction is expected to be dominated by the non-magnetic and magnetic zero frequency contributions [for more detail see Sec.\,(S.7) in the Supplementary Information]. However, it is important to emphasize that when the interaction occurs in a fluid the excited state lifetime will be significantly reduced. This will impact the potential existence of excited interactions in a fluid at large separations. 

%\section{Material modeling}

%\subsubsection{Modeling magnetic fluid}

The nanoparticles we consider have large enough dimensions to approximate their effective dielectric functions with those of bulk materials. The dielectric functions at imaginary Matsubara frequencies are obtained through the Kramers--Kronig relation
\begin{equation}
\varepsilon(i \xi_p) = 1+\frac{2}{\pi}\int_0^\infty d\omega\, \frac{ \omega\, \varepsilon''(\omega)}{\omega^2+\xi_p^2}.
\label{eq:KramKronEq}
\end{equation}
%
%\textcolor{pink}{We use available experimental data from the literature for polystyrene (PS)\,\cite{Zwol1}  nanoparticles. This reference is also used to provide the dielectric function for Toluene. For ZnO\,\cite{Bostrom2023_PhysRevB.108.125434} we use results from density functional theory (DFT) that, together with additional materials information, are described in detail in Sec.\,(S.5) of the Supplementary Information. }
%\textcolor{red}{Clas' thoughts: I don't think "PS nanoparticles" is correct. Even if we describe them as nanorods, I guess the reference does not present data for PS nanorods (instead, the geometry is part of the theoretical model). Moreover, PS is well used, so no need to define again. I instead suggest the following lines, and I have also changed the text about ZnO, gold, and magnetite according to reviewer's comments (and data for ZnO is not from a reference; it is new calculations for this article.). I have also modified the two lines "We use experimental data from the literature for toluene 32 at T = 298.15 K. Gold coatings are taken to have bulk properties given in the literature." }
The geometries of the nanoparticles and nanorods are described within our theoretical model, whereas the material properties are  macroscopic attributes. 
We use available room-temperature experimental data  for PS as provided by van Zwol, {\it et al.}.\,\cite{Zwol1}. 
For ZnO we calculate the imaginary part of the dielectric function within the density functional theory (DFT) as implemented in the Vienna Ab initio Simulation Package.\,\cite{VASP1999} To correctly mimic the experimental band-gap energy of 3.4\,eV, we employ the hybrid functional with 37.5\% Fock exact exchange, which is a widely adopted parameter choice for this oxide. The $GW$-type pseudopotentials ensure better accuracy in describing high-frequency optical transitions. Convergence with respect to $k$-space mesh is considered.\,\cite{Crovetto2016}
Further computational details, together with additional materials information, are described in Sec.\,(S.5) of the Supplementary Information.

The optical properties of the magnetic fluid are estimated by taking into account the volume fraction of commercially available gold-coated magnetite particles in a toluene mixture. We use the mixing model presented by 
Chen {\it et al.}\,\cite{ChenEffPermiLayered1998} as,
% we use the notation in Fig.\,\ref{schematicsEffectiveMedia},

\begin{equation}
    \varepsilon_3=\frac{\varepsilon_{3,3}(1+2 \Lambda_3)}{1-\Lambda_3}; \quad \Lambda_3=(\phi_{1}+\phi_{2})\frac{\varepsilon_{s}-\varepsilon_{3,3}}{\varepsilon_{s}+2\varepsilon_{3,3}},
    \label{2mixturedielEqA}
\end{equation}
where the effective dielectric function of the sphere ($\varepsilon_{s}$) is
\begin{equation}
    \varepsilon_s=\frac{\varepsilon_{3,1}(1-2 G)}{1+2G}; \quad G=\frac{\varepsilon_{3,1}-\varepsilon_{3,2}}{2 \varepsilon_{3,1}+\varepsilon_{3,2}}\frac{\phi_{2}}{\phi_{1}+\phi_{2}}.
    \label{2mixturedielEqB}
\end{equation}
Here, $\phi_i$ and $\varepsilon_{3,i}$ are the volume fraction and dielectric function of  toluene ($i=3$), magnetite ($i=2$) and gold ($i=1$) for modeling the magnetic fluid where $D$ is the diameter of the magnetite core and the thickness $R_{\rm Au}$ of the gold coating, as illustrated in Fig.\,(\ref{fig:scheme}).  The radius of the gold coated nanoparticles is defined as $R_1 = D/2 +  R_{\rm Au}$.\\
In the current work, we consider the case where the amount of magnetite is 5\% and 7\% ($\phi_2=0.05, 0.07$). The corresponding volume fraction of the gold coating is therefore given by $\phi_1=\phi_2 \times[(2 R_1/D)^3-1]$. We consider four typical cases where the average diameter ($D$) of the magnetite nanoparticles is 10, 13, 15 and 17 nm. We use experimental data for toluene at $T=298.15$\,K from the literature.\,\cite{Zwol1}  
The dielectric functions of magnetite and gold are from 
DFT calculations in our recent work,\,\cite{Carretero_PRB_2025_PhysRevB.111.085407} where room-temperature thermal electronic excitation was considered for the small-gap magnetite compound. In the main text, we report some important results that are obtained for pure magnetite particles (i.e. for $R_1=D/2$). Most of the results obtained using gold-coating are included in Sec.\,(S.8) of the Supplementary Information. The size of the magnetite nanoparticles and gold-coated magnetite nanoparticle can be adjusted to control trapping behavior for pairs of unequal nanorods.

The permeability of different  ferrofluids was described by Klimchitskaya {\it et al.}\,\cite{klimchitskaya2019impact,VelichkoKlimchitskaya2020}, where 
 numerical calculations were presented for a water-based ferrofluid containing a 5 \% volume fraction of  10 nm diameter magnetite nanoparticles. The fluid was enclosed between two dissimilar plates, Au and SiO$_2$. 
 It was found that the magnetic particles caused the CL force to be repulsive for all the gap widths considered, between 100 nm and 2 $\mu$m.  Both the Drude and the plasma dispersive models for the Au plate were calculated, and no found difference.
These results are of obvious technological interest. For microdevices  the stiction effect  may be a serious difficulty, and the magnetically tuned repulsiveness may therefore  be helpful.
\begin{figure}[!ht]
    \centering
    \includegraphics[width=0.7\linewidth]{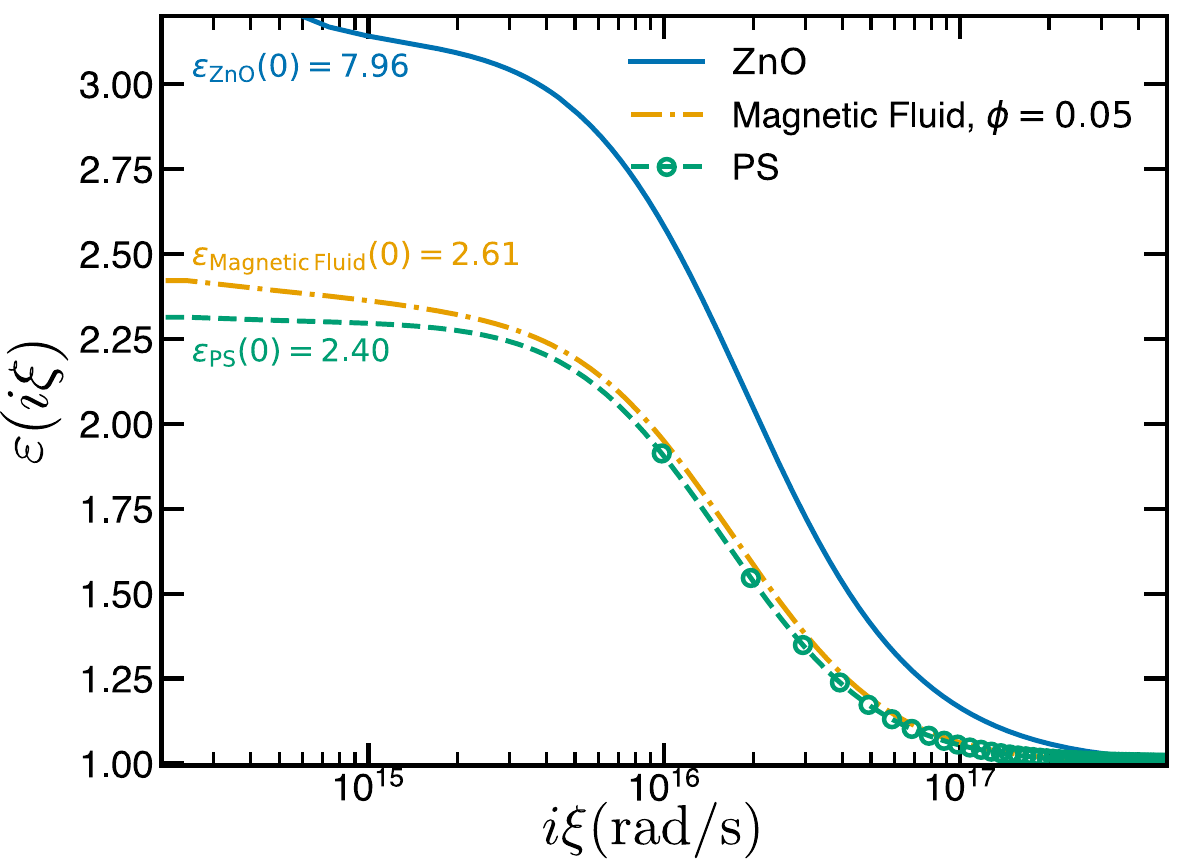}
    \caption{(Color online): 
The dielectric functions on the imaginary axis as functions of frequency $\xi$. The parametrized data for PS and toluene are based on optical experiments given by van Zwol {\it et al.}\,\cite{Zwol1}. 
The magnetic fluid consists of a mixture of toluene and magnetite particles, using  Eq.\,(\ref{2mixturedielEqA}) for a volume fraction of $\phi = 5\%$. The response functions of gold and magnetite are DFT results presented in Ref.\,\cite{Carretero_PRB_2025_PhysRevB.111.085407},  whereas the data for ZnO are from present DFT calculations.
}  
    \label{dielectric_phi_0.05}
\end{figure}

For water (close to being a non-magnetic fluid) with a fraction $\phi$ of magnetite Klimchitskaya {\it et al.} estimated the permeability to be\,\cite{klimchitskaya2019impact,VelichkoKlimchitskaya2020}
\begin{equation}
    \mu_3(0)=1+\frac{2\pi^2\phi}{9}\frac{M_s^2 D^3}{k_{B} T}.
    \label{mu3}
\end{equation}
Toluene is only a little more magnetic ($\mu=0.99996\approx1.00$) so within the level of accuracy in our work the permeability of both water and toluene-based mixtures can be estimated with the approximation used by Klimchitskaya and Mostepanenko for water-based mixtures\,\cite{klimchitskaya2019impact,VelichkoKlimchitskaya2020}. 
% \Jeet{I don't think it is important: 
% %Then, irrespective of the specific non-magnetic carrier liquid, and 
% If we for example assume a magnetite volume fraction of $\phi=$0.05, $\mu_3(0)$=1.24 and 2.9 for magnetite nanoparticles with $D$ =10 and 20\,nm diameter, respectively.}
Using Eq.\,(\ref{mu3}) we derive the permeability for a range of  volume fractions (from 0.05 to 0.07) and diameters (from 10 to 17\,nm). The cylindrical nanoparticles we consider are non-magnetic ($\mu_1(0)\approx \mu_2(0)\approx1$).

% \mb{EDIT THE TEXT BELOW WHICH IS DIRECTLY COPIED BY ME/Mathias FROM THE PRL Manuscript: Add description of dielectric function of toluene and magnetite and then edit the text below about mixing models}

% \mb{To describe the effect of the volume fraction of
%     magnetite nanoparticles we use a Lorentz-Lorenz-like model for the effective
%     dielectric function\,\cite{Aspnes,Krag_LorenzLorentzFormula},
    
%     \begin{equation}
%         \varepsilon_{3}=\frac{1+2 \Lambda_{3}}{1-\Lambda_{3}}; \quad \Lambda_{3}=
%         \sum_{\nu=1,2}\Phi_{\nu}\frac{\varepsilon_{3,\nu}-1}{\varepsilon_{3,\nu}+2}
%         , \label{3mixturediel}
%     \end{equation}
%     where $\Phi_{\nu}$ and $\varepsilon_{3,\nu}$ are the volume fraction and dielectric
%     function of the toluene ($\nu=1$) and magnetite ($\nu=2$). One notes that the Lorentz-Lorentz model relies on the quasi-static approximation for dilute media \,\cite{Aspnes,Krag_LorenzLorentzFormula}. Its use in related systems has been successfully tested theoretically~\cite{klimchitskaya2019impact,VelichkoKlimchitskayaNepomnyashchaya2020} and experimentally~\cite{ZhangNature2024}.
%         Furthermore,  different mixing models for the effective dielectric functions of multi-component fluids give similar results for the attraction-repulsion transitions in the considered system  \,\cite{Markel_JOptSocAmA2016}.

% }

\begin{figure}[!ht]
    \centering
    \includegraphics[width=0.85\linewidth]{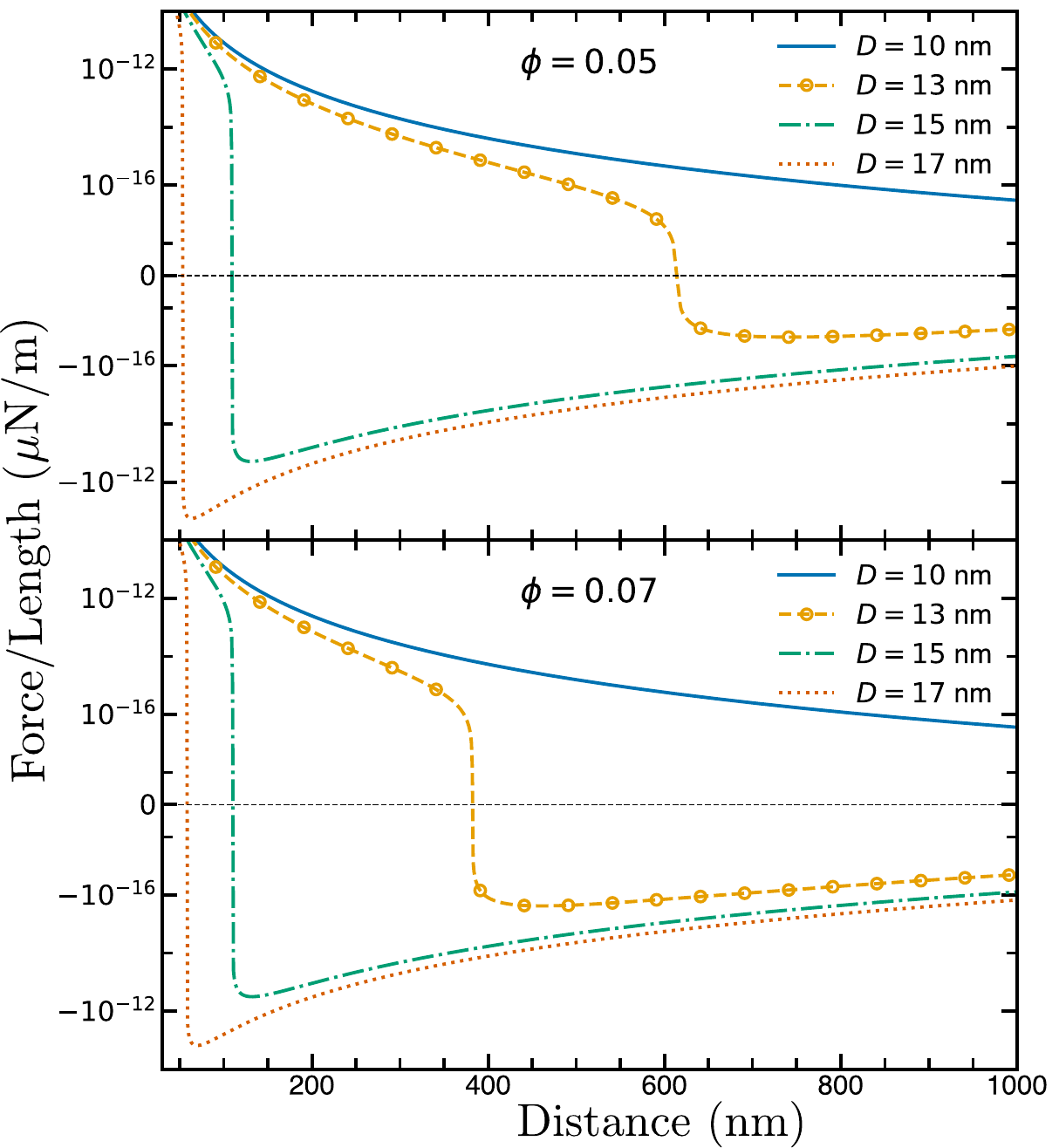}
    \caption{(Color online:) The CL force per unit length between two cylindrical objects (ZnO and PS respectively) with the same radius $a=20$ nm as a function of center-to-center distance $R$ when immersed in a ferrofluid with  5\% and 7\% concentrations of magnetite nanoparticles of diameter 10, 13, 15 and 17 nm. }
    \label{fig:force_vs_distance_phi_0.05}
\end{figure}

We recently investigated the interaction
between a planar PTFE surface and a spherical PS surface in the presence of a magnetic fluid\,\cite{pal2025trapping}.
In this case both the non-retarded and retarded interaction could be tuned to be repulsive while zero frequency contributions to the CL interaction (both TM and TE) where attractive. This led us to conclude that it is possible to have a {\it magnetic contribution} to CL trapping\,\cite{pal2025trapping}. However, trapping still occurred when the magnetic effects were ignored (i.e. the magnetic permeability set to one).

\begin{figure}[!ht]
    \centering
    \includegraphics[width=0.8\linewidth]{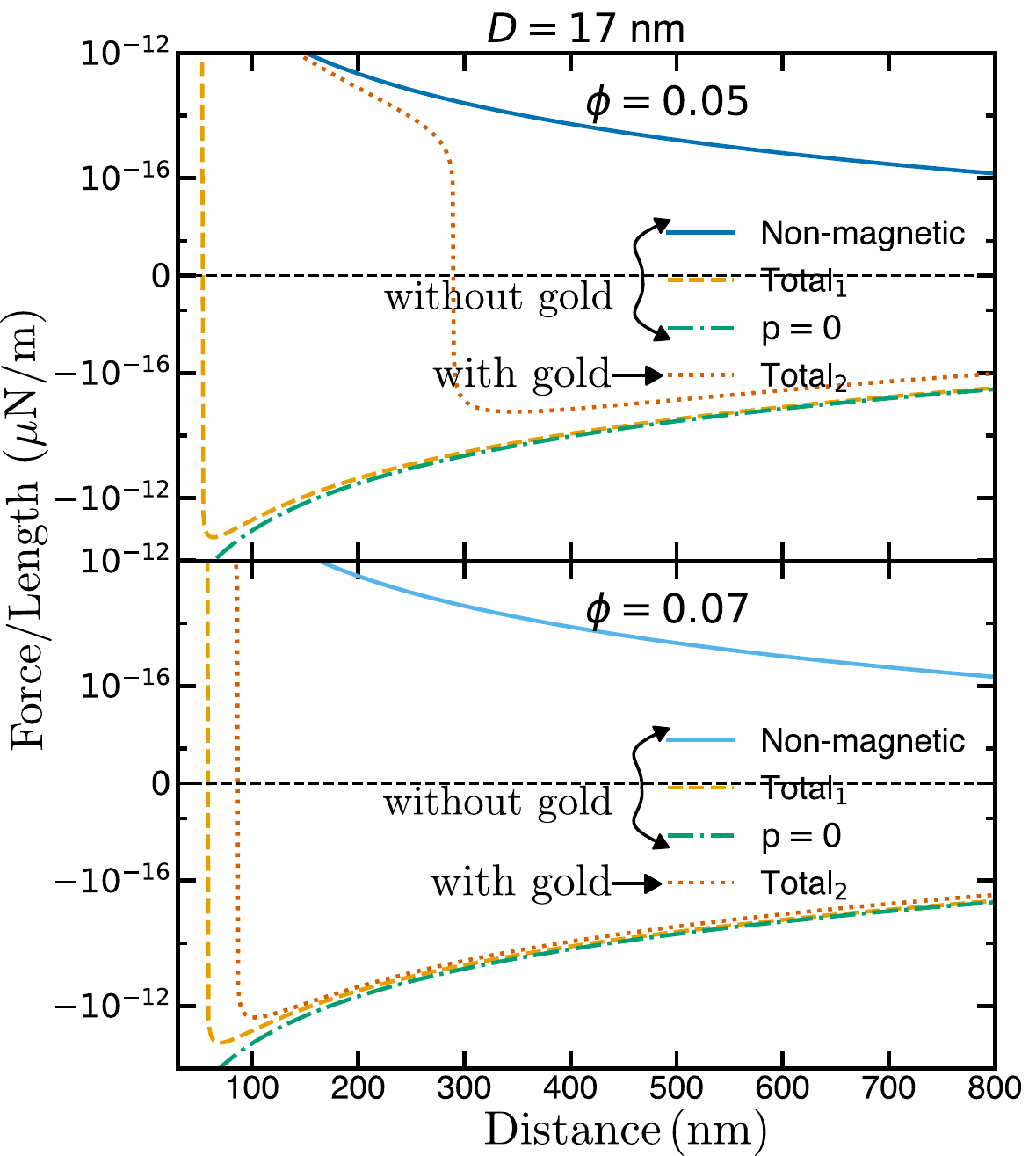}
    \caption{(Color online): The CL force per unit length between two ZnO and PS cylindrical objects was evaluated for  volume fraction $\phi = 5\%, 7\%$ and a nanoparticle diameter $D= 17$ nm. The terms ``without gold'' and ``with gold'' refer to magnetite nanoparticles  without and with a gold coating, respectively.  For gold coating, layer thickness was taken to be, $R_{\rm Au} = 2$ nm. Total$_{1}$ is the fully retarded calculation for all frequencies without gold coating whereas Total$_{2}$ reflects the retarded  force with gold coating and $p=0$ denotes the zero frequency contribution.} 
    \label{fig:force_vs_distance_phi_0.05_goldcoating}
\end{figure}

We will in the current work explore a more versatile system consisting of nanorod pair complexes where the materials are described by the dielectric functions shown in Fig.\,(\ref{dielectric_phi_0.05}). Here the intermediate magnetic fluid has a dielectric function in between those of ZnO and PS. This indicates that the non-retarded and retarded CL interaction between a ZnO and PS nanorod will be repulsive\,\cite{Dzya}. 

\begin{figure}[!h]
    \centering
    \includegraphics[width=0.7\linewidth]{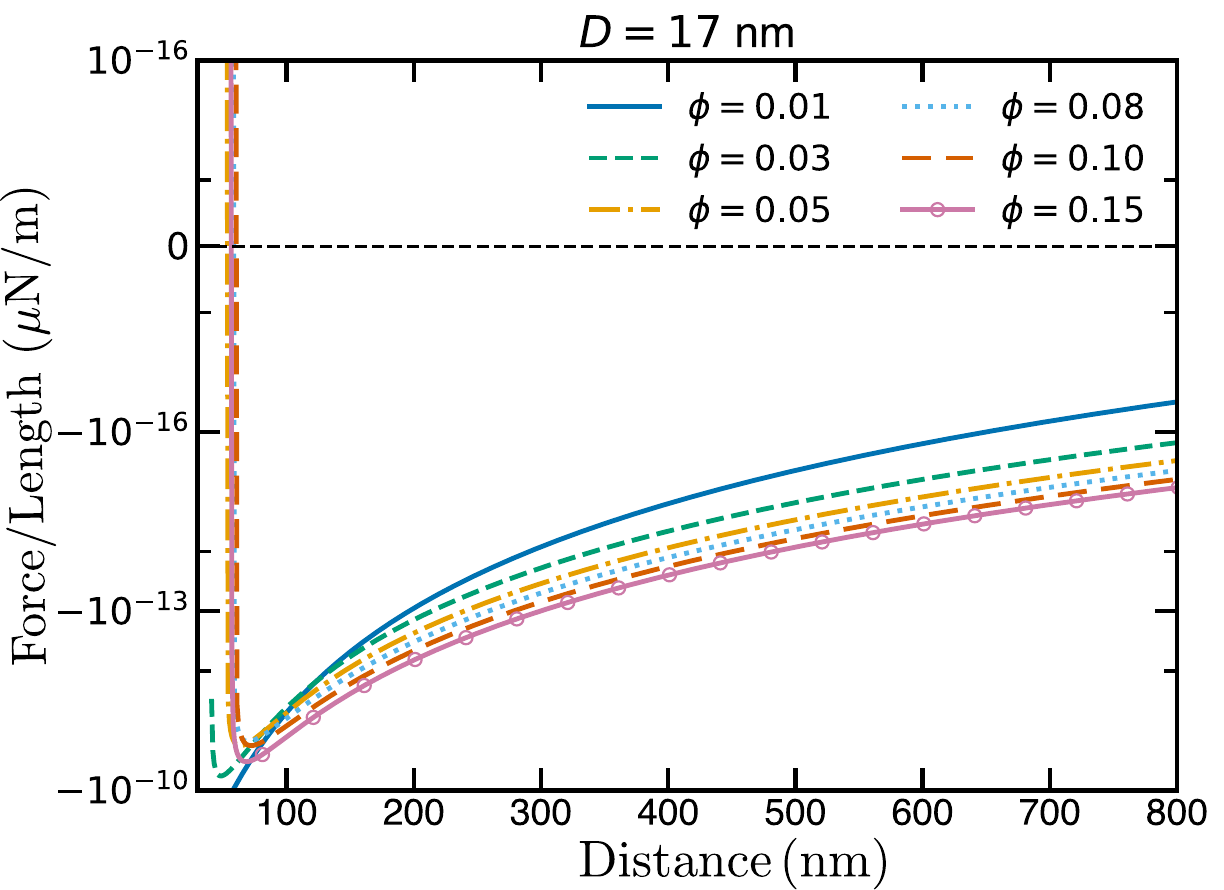}
    \caption{(Color online:) The CL force per unit length between two cylindrical objects (ZnO and PS respectively) with the same radius $a=20$ nm as a function of center-to-center distance $R$ when immersed in a ferrofluid with different concentrations of ($\phi$=0.01-0.15) magnetite nanoparticles of diameter 17\,nm.}
    \label{fig:varphi}
\end{figure}

In Fig.\,(\ref{fig:force_vs_distance_phi_0.05}) we show that in the case of a pair of ZnO and PS nanorods in a magnetic fluid with toluene mixed with 5-7\%  magnetite it is possible to obtain trapping that is {\it entirely} controlled by the magnetic contribution to the CL force. When the magnetic permeability of the fluid is set to one ($\mu_3 = 1$), the system only experiences repulsion. This dependence of trapping on the magnetic permeability has not been possible to achieve in the past. Notably, when the diameter of the magnetite nanoparticles decreases with fixed volume fraction of magnetite, or a weak magnetic field is applied, leading to reduced magnetic permeability of the fluid\,\cite{ZhangNature2024}, the trapping occurs at increasingly larger nanorod pair separations. The trapping can also be controlled by keeping the diameter constant and varying the volume fraction. As illustrated in Fig.\,(\ref{fig:varphi}), the trend is that larger $\phi$ shifts the force minimum towards larger separations and making it more shallow. Ultimately, in the absence of magnetic permeability effects or very small volume fraction, there is no trapping at all and the nanorod complex is split into two separate nanorods.
Replacing the magnetite nanoparticles with commercially available gold-coated magnetite nanoparticles leads to an additional mechanism to tune the trapping distance as can be seen in Fig.\,(\ref{fig:force_vs_distance_phi_0.05_goldcoating}). With changes in the gold coating the dielectric function of the background media changes and the trapping distance can be shifted. 

% \mb{This leads towards a new method for tuning of forces between unequal nanorods from attractive potentials onto repulsive-attractive trapping potentials to systems which have pure repulsion. Hence, with clever manipulations, we can have complete adhesion, trapped tunable clusters, or fully destabilized clusters. Notably, this opens up a versatile toolbox for  engineering of nanoclusters in magnetic fluids, relevant for NEMS and MEMS, via tubable magnetic Casimir forces.
% Here, we stress that the proof-of principle theory developed here requires the trapping to take place at separations large compared to the size of the nanorods. This means the trapping minima are shallow and susceptible to destabilization due to thermal fluctuations (see Supplementary Information). This can be overcome using proximity force theorem which is valid in the opposite limit (short separations compared to the size of the nanorods) where the trapping can be made more stable.   In the long-term perspective, with careful selection of materials, the basic ideas proposed here could potentially be extended to the controlled complexation of specific biopolymers with artificial nanothreads via added magnetic nanoparticles in biofluids.}

 Non-magnetic Casimir forces have been exploited, both theoretically and experimentally, to produce Casimir trapping and non-adhesion, via repulsive forces, in NEMS and MEMS\cite{Munday2009,esteso2015nanolevitation,WoodsRevModPhys.88.045003,zhao2019stable,Woods2020,esteso2022effect,MundayReview2024,WilliamsonJPCL2025}. We have focused in the current work on the novel results where only the zero-frequency TE term changes, as trapping occurs over a range of different $\phi$ values. The repulsion occurs when $\phi$ is tuned such that the dielectric function for the fluid is between the values of the dielectric functions of the interacting nanorods for a wide range of frequencies. Our work demonstrates that the trapping can be controlled by both changes in the size of the magnetic nanoparticles and the concentration in the fluid. The exact values for these quantities can be adjusted to achieve trapping for different separations and for different material combinations. The proposed framework suggests a novel route for actively tuning the interacting forces between two unequal nanorods  where a purely attractive potential can be driven into repulsive–attractive trapping potentials and even into regimes of complete repulsion. By a careful choice of the material parameters, one can tailor finite nanosystems that give rise to reversible adhesive behavior, stable tunable trapping and on-demand destabilization of clusters. These results demonstrate a general platform for nanocluster design in magnetic colloids where adjustable magnetic Casimir forces can be exploited for NEMS/MEMS applications.

Note that the proof-of-principle theory presented here assumes that trapping takes place at separations much larger than the diameter of the nanorod. In this situation, the trapping minima are very weak and can be affected by thermal noise [see Sec.\,(S.6) in the Supplementary Information]. Nevertheless, this restriction can be mitigated by applying the proximity force approximation (PFA), valid in the short-distance regime, where enhanced stability of the trapping potential is expected.

 Going forward, the framework we have developed here indicates some exciting prospects for future extension of tunable dispersion-force control{\color{blue}{\,\cite{Munday2009,esteso2015nanolevitation,WoodsRevModPhys.88.045003,zhao2019stable,Woods2020,esteso2022effect,MundayReview2024,WilliamsonJPCL2025}}} to biomolecular materials. In particular, these ideas could be extended to direct biopolymer complexation with artificial nanothreads via magnetic nanoparticles in biofluids with a suitable choice of materials. This has a great potential in a variety of future diagnostic and therapeutic nanomedicine research\,\cite{CoeneMagneticNPJApplPhys2022}. It also provides a route to field-controlled assembly in hybrid soft–magnetic nanostructures and could also find utility for biosensing, targeted delivery, reconfigurable nanoarchitectures, and beyond.

%Systems with stable repulsion or stable attraction between two unequal, elongated nanoparticles will be considered first.

% \begin{figure}[!h]
%     \centering
%     \includegraphics[width=\linewidth]{try_2.pdf}
%     \caption{(Color online:) The Casimir force per unit length between two cylindrical objects (ZnO and PS respectively) with the same radius $a=20$ nm as a function of center-to-center distance $R$ when immersed in a ferrofluid with  5\% and 7\% concentrations of magnetite nanoparticles of diameter 10, 13, 15 and 17 nm. }
%     \label{fig:force_vs_distance_phi_0.05}
% \end{figure}

%{\bf{FIGURES: find systems with only repulsion and only attraction and sign change with full theory (that is mu3 different from 1).}}

%\subsection{Casimir Trapping controlled by permeability of toluene-magnetite fluid mixture}

%PS-fluid-PTFE

%{\bf{FIGURES: Phi=0, 0.005, 0.01 using $R_2=R_1$ and D=2 nm, 5nm, and 10 nm. a=50nm. With and without including mu different from 1.}}

%  \begin{acknowledgments}
\vspace{0.3in}
\noindent \textbf{\Large Data availability}:\,The code and data that support the findings of this study are available in \href{https://github.com/subhojit-pal/Cylindrical-nanorods-with-a-background-magnetic-fluid-}{Github}.

  \section{Acknowledgement}
 SP and UDG acknowledge support from the Marie Sk{\l}odowska-Curie Doctoral Network TIMES, grant No. 101118915 and SPARKLE grant No. 101169225.
  MB's contributions to this research are part of the project No. 2022/47/P/ST3/01236 co-funded by the National Science Centre and the European Union's Horizon 2020 research and innovation programme under the Marie Sk{\l}odowska-Curie grant agreement No. 945339. 
  Institutional and infrastructural support for the ENSEMBLE3 Centre of Excellence was provided through the ENSEMBLE3 project (MAB/2020/14) delivered within the Foundation for Polish Science International Research Agenda Programme and co-financed by the European Regional Development Fund and the Horizon 2020 Teaming for Excellence initiative (Grant Agreement No. 857543), as well as the Ministry of Education and Science initiative “Support for Centres of Excellence in Poland under Horizon 2020” (MEiN/2023/DIR/3797).
  We gratefully acknowledge Poland's high-performance computing infrastructure PLGrid (HPC Centers: ACK Cyfronet AGH) for providing computer facilities and support within computational grant no. PLG/2023/016228 and for awarding this project access to the LUMI supercomputer, owned by the EuroHPC Joint Undertaking, hosted by CSC (Finland) and the LUMI consortium through grant no. PLL/2023/4/016319. CP acknowledges access to high-performance computing resources from NAISS, provided by NSC and PDC. LMW acknowledges financial support from the US Department of Energy under Grant No. DE-FG02-06ER46297.
% \end{acknowledgments}

% \begin{suppinfo}
% \label{SI}

% \end{suppinfo}

% \bibliographystyle{achemso}
% \newpage

\bibliography{jeet}

% \bibliography{GM}

\end{document}